\documentclass[letterpaper]{article} 
\usepackage[preprint]{paperstyle}
\usepackage[hyphens]{url}  
\usepackage{graphicx} 
\usepackage{natbib}  
\usepackage{caption} 
\usepackage{booktabs}
\usepackage{amsmath}
\usepackage{amssymb}
\title{Listen, Do Not Copy: Internalizing Audio-Grounded Scaffold Context\\ for Robust Omni-Model Speech Understanding}
\author{
    Pengfei Zhang\textsuperscript{\rm 1},
    Biao Tian\textsuperscript{\rm 2},
    Tianxin Xie\textsuperscript{\rm 1},
    Minghao Yang\textsuperscript{\rm 1},
    Xiangang Li\textsuperscript{\rm 2},
    Li Liu\textsuperscript{\rm 1}\corresponding
}
\affiliations{
    \textsuperscript{\rm 1}AI Thrust, Information Hub, The Hong Kong University of Science and Technology (Guangzhou)\\
    \textsuperscript{\rm 2}Token Foundry, Alibaba Group, China\\
    avrillliu@hkust-gz.edu.cn
}

\begin{document}

\maketitle

\begin{abstract}
Omni models transcribe clean, single-speaker speech well, but their accuracy drops sharply when speakers overlap and the scene is noisy, exactly where knowing who said what matters most. A natural fix is a short scene description. We show why this is risky: answer-bearing text lets the model copy instead of listen, so the score rises although nothing has been heard; a silent test exposes this shortcut at once. We call this failure mode \textbf{perception bypass} and address it with \textbf{Audio-Grounded Scaffold Context (AGSC)}. AGSC links three steps: first, we build clues from audio to guide listening without giving the answer; second, answer-overlap and silence tests probe them for leakage and audio dependence; finally, those clues scaffold training but vanish at test time, yielding no-clue capability. Across three heterogeneous Omni models, training on AGSC lowers no-clue capped mean permutation word error rate (mpWER) on overlapping, noisy speech from $25\%$--$71\%$ to $9\%$--$15\%$. For streaming control, we train with GDPO under separately normalized format, gate, and transcript rewards: a gate-only stage learns when a clue is worth using, and a full-chain stage jointly trains the scene gate and the speaker-attributed transcript. After internalization, AGSC adds almost no inference overhead.
\end{abstract}

\section{Introduction}
Imagine a meeting recording where two people briefly speak at once over the hum of an air conditioner. An Omni model is a multimodal large language model that handles text, images, audio, and video in one end-to-end system~\citep{qwen3omni,minicpmo,mingflash}. It may transcribe the clean parts well yet fail at the overlap, merging voices, dropping a speaker, or treating noise as words. Such conditions are common in meetings, call centers, and interviews, exactly where knowing who said what matters most.

Textual scene descriptions seem like a natural aid when audio is unclear: supplying a full transcript as context raised the mean accuracy of three Omni models from about $53\%$ to $94\%$. The jump was too good to be true: on the primary-transcript task, the two stronger models copied a wrong contextual answer $94\%$--$99.8\%$ of the time, and with silence in place of audio all three still reached $100\%$ accuracy. We call this failure mode \textbf{perception bypass} and turn it into a pass-or-fail test for context quality.

That failure suggests a hypothesis: Omni models already hold useful acoustic capabilities that complex scenes leave underused. The premise matches how such models are trained: Qwen3-Omni pretrains its audio encoder on 20 million hours of audio~\citep{qwen3omni}, large-scale diverse supervision is shown to improve noise robustness~\citep{whisper}, and pretraining that injects overlapped, noisy speech improves speaker tasks~\citep{wavlm}. Our results agree: with no training, an answer-screened prefix clue cuts hard-slice error by up to $18$ points (Figure~\ref{fig:streammat}). Yet a context that hands over words instead of pointing at the audio falls straight back into perception bypass. Audio-Grounded Scaffold Context (AGSC) resolves this tension by construction: its clues expose only limited scene structure, pointing without answering, and a silent-audio control checks that no clue succeeds without sound.

\begin{figure*}[t]
\centering
\includegraphics[width=\textwidth]{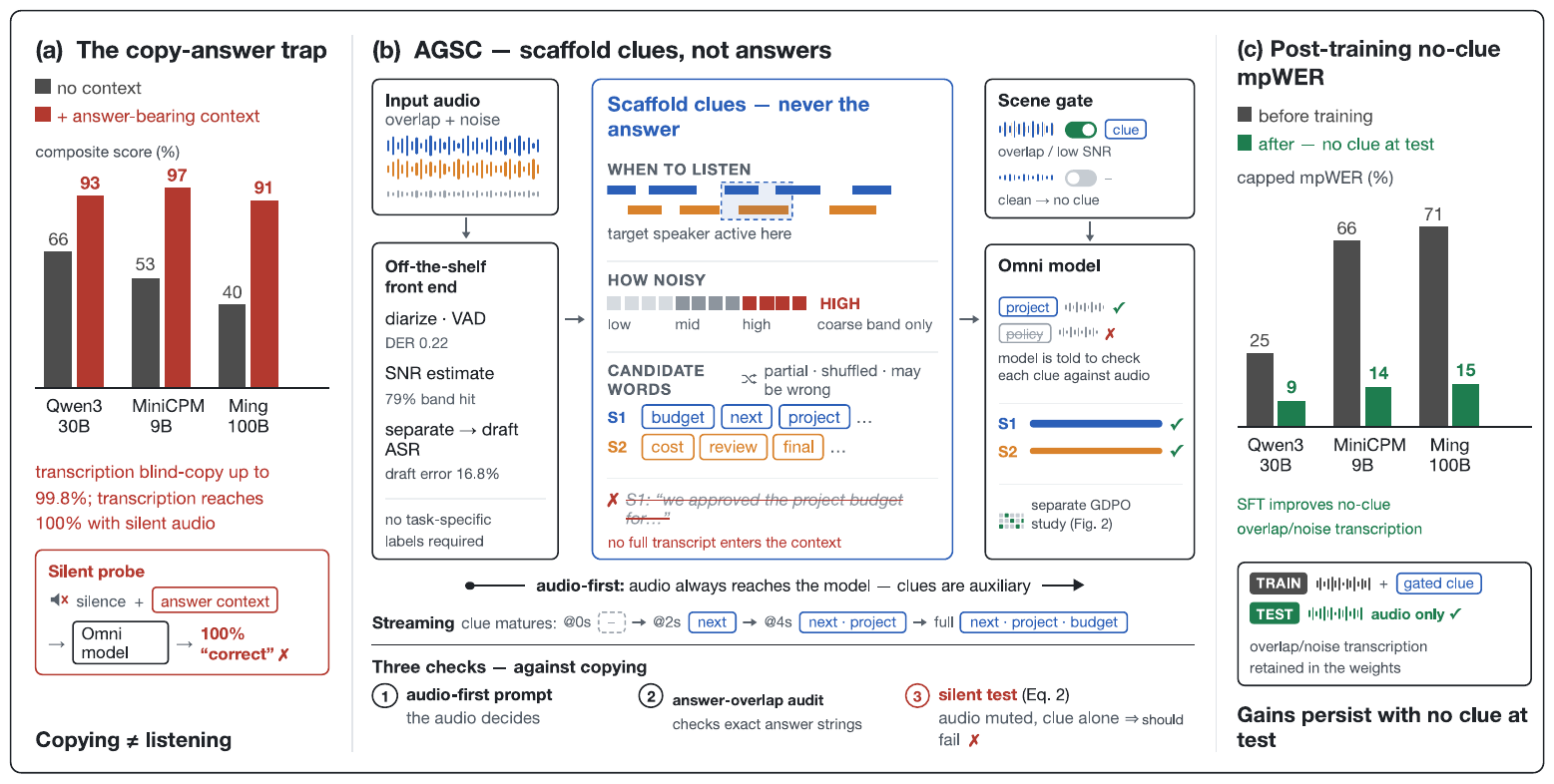}
\caption{Overview. (a) Answer-bearing context inflates scores by copying. (b) AGSC provides limited-content scaffold clues with three safeguards. (c) After supervised training, capped mean permutation word error rate (mpWER) stays lower when the clue is removed. The \emph{silent test} is an aggregate control; we use the no-clue gain as operational evidence of \emph{internalization}.}
\label{fig:overview}
\end{figure*}

Rather than telling the model what was said, AGSC supplies three scaffolds (Figure~\ref{fig:overview}): target-speaker activity, scene noise level, and partial, shuffled words, meant to guide listening, not to answer in its place. Three safeguards limit or expose context-only shortcuts: an \textbf{audio-first instruction}, an \textbf{answer-overlap screen}, and a \textbf{silent-audio control}.

Two training paths move this ability out of the prompt and into the model. Supervised fine-tuning (SFT) uses gated clues as temporary scaffolds: after training we remove them entirely. If the gain survives with no clue, the skill lives in the weights rather than in the prompt; this surviving no-clue gain is what we call \textbf{internalization}. Group reward-decoupled normalization policy optimization (GDPO)~\citep{gdpo} governs streaming, where a clue can still pay off at run time: with separate format, gate, and transcript rewards, gate-only training learns when a clue is worth injecting, and full-chain training learns gate and speaker-attributed transcript jointly.

This paper makes three contributions. \textbf{(1)} We identify perception bypass and turn silent audio into a design test for context. \textbf{(2)} We introduce AGSC (an automatic synthesis pipeline) and \textbf{Context-Speech Bench (CSB)}, covering overlap, noise, meetings, streaming clues, and gating. \textbf{(3)} We show internalization: supervised training improves no-clue transcription, and one GDPO stage trains gate and transcript jointly. Code, manifests, evidence, and public-corpus audio recipes are released with the paper.

\section{Related Work}
Audio-language models join an audio encoder with a large language model, ranging from Qwen-Audio~\citep{qwenaudio} and SALMONN~\citep{salmonn} to Omni systems that unify perception and generation in one system. The benchmarks that track this progress, AIR-Bench, MMAU, and Dynamic-SUPERB, measure broad listening and reasoning ability, and some report text-only or cascade baselines~\citep{airbench,mmau,dynamicsuperb}; yet to our knowledge none tests one audio model with an auxiliary text clue on both real and silent audio, exactly where perception bypass would surface. We add this control to Qwen3-Omni-30B-A3B-Instruct~\citep{qwen3omni}, MiniCPM-o 4.5~\citep{minicpmo}, and Ming-flash-omni-2.0~\citep{mingflash}.

For the task itself, one route chains diarization~\citep{pyannote}, voice activity detection~\citep{silero}, automatic speech recognition (ASR)~\citep{whisper}, and sometimes separation~\citep{sepformer}, where early errors propagate; another trains one model to emit the words of several overlapping speakers~\citep{tsot}, uses estimated speaker profiles~\citep{tsvad}, or targets distant, noisy meetings~\citep{chime8}. Both lines treat the front end or the model itself as the answer path. We instead demote the same front-end tools to clue producers: their outputs may only point the model at the audio, must survive the answer-overlap screen, and must earn their place by measurement. The backbone stays frozen; only low-rank adaptation (LoRA) modules on the attention projections are trained~\citep{lora}.

A separate tradition supplies models with trusted text: retrieval-augmented generation draws on an external memory~\citep{rag}, chain-of-thought prompting exposes intermediate reasoning~\citep{cot}, and contextual biasing hands a recognizer reliable word lists such as contact names~\citep{clas}. Text-only correctors go further and rewrite recognition or diarization outputs with a language model~\citep{hyporadise,diarizationlm}, but cannot consult the audio to fix a fluent yet wrong draft. Our context inverts their shared trust assumption: its candidate words are automatic, possibly wrong, partial, and shuffled, so the audio, not the text, has the final say.

The risk motivating this inversion is well documented. Perception bypass is a form of shortcut learning~\citep{shortcut}; our silent test parallels the language-prior controls that rebalance visual question answering~\citep{vqa2}; and over-trust of one modality against another is reported for text over vision~\citep{blindfaith} and visual or language priors over audio evidence~\citep{audiohalluc,cmm}. Closest to us, recent work shows that large audio models can answer from text alone and uses that signal to split training data~\citep{audiocontrib}. We push the observation further and make silence a design criterion that every clue must fail.

Two ideas frame our training questions. Context distillation moves the benefit of instructions or scratchpads into model weights~\citep{ctxdistill}; we ask its audio counterpart: does clue-conditioned training survive clue removal? IRAF learns a continuous gate that suppresses an unreliable input before fusion in full-duplex dialogue~\citep{iraf}; our discrete gate instead decides whether a limited-content text clue enters while every speaker is kept.

\section{Method}
AGSC has three linked steps: it synthesizes answer-incomplete clues from audio, screens them for direct answer exposure and audio dependence, and uses them as removable training scaffolds. A separate GDPO study learns a joint gate-plus-transcript output for streaming control.

\subsection{Problem and Perception Bypass}
\label{ssec:trap}
Given an audio clip $a$ with $K$ speakers, the model must output one line per speaker. Let $y=(y_1,\dots,y_K)$ be the gold transcriptions and $\hat{y}$ the predicted lines. We score outputs with mean permutation word error rate (mpWER), a speaker-averaged variant of concatenated minimum-permutation word error rate (cpWER)~\citep{cpwer}. It is the equally weighted mean of per-speaker word error rates (WER) under the best speaker permutation (lower is better):
\begin{equation}
\mathrm{mpWER}(y,\hat{y})=\min_{\rho\in S_K}\ \frac{1}{K}\sum_{k=1}^{K}
\frac{\mathrm{ED}\big(y_k,\hat{y}_{\rho(k)}\big)}{|y_k|}.
\label{eq:mpwer}
\end{equation}
In Equation~(\ref{eq:mpwer}), $\rho\in S_K$ assigns lines to reference speakers, $\mathrm{ED}$ is word-level edit distance, and $|y_k|$ is reference length. Missing output lines are padded as empty. For this two-speaker benchmark, including streaming evaluation, we score only the first two output lines. The permutation is over anonymous output lines: mpWER measures attribution of text to separate speaker streams, not recovery of speaker identities. Let $c$ be a textual context. We seek a context whose gain depends on $a$, not on $c$ alone.

An overly informative $c$ can make $a$ irrelevant; this is perception bypass. The introduction showed that a transcript-bearing context inflates accuracy by copying: the answer is a copyable substring, wrong labels are copied, and silent audio still scores perfectly. We turn this finding into a design criterion. For model $M$, a clue $c$ is desirable for sample $(a,y)$ only when it cannot solve the sample without audio:
\begin{equation}
\mathrm{mpWER}\big(y,\,M(a_{\varnothing},c)\big)\ \ge\ \eta,
\label{eq:silent}
\end{equation}
where $a_{\varnothing}$ is a silent track of the same length as $a$. We set $\eta=1$ (an mpWER of at least $1$) for transcription; for classification probes, we report accuracy and use chance level as the analogous boundary. Equation~(\ref{eq:silent}) states the desired sample-level property; no per-sample filtering is applied. At evaluation, we apply the same idea at the aggregate level by measuring performance after the audio is silenced. This is an empirical control, not a proof that every shortcut is absent.

\subsection{AGSC: Clues, Not Answers}
\label{ssec:agsc}
The silent criterion says what a clue must not contain, not what makes one helpful. Our working principle is that only a high-quality scaffold assists a complex scene: wrong content can mislead the attention it should direct, and a clue offered to an already clean, single-speaker segment has little to gain and may instead distract. AGSC therefore controls both what a clue says and whether one should exist at all.

What a clue says matches the diagnosed uncertainty. For overlap, where the ambiguity is \emph{who} and \emph{when}, AGSC provides a time window per speaker, such as \emph{target active 0.00--3.85\,s, female}, but no words. For single-speaker noise, where the listening condition is unknown, it provides a coarse noise level, speech regions, and a partial, shuffled word list that keeps about two thirds of the candidate words and removes exact answers such as digits. For overlap plus noise, a separator produces a rough draft for each voice; AGSC keeps half of the content words, adds up to four distractors from other recordings, removes duplicates, and shuffles the result; the full draft is never shown (Figure~\ref{fig:overview}b). Because each clue is shuffled, partial, and free of exact values, it points at the audio instead of answering.

Because every clue is machine-made and imperfect, three safeguards limit or expose shortcuts. The audio-first instruction marks each clue as automatic and possibly wrong. The answer-overlap screen rejects directly copyable answer strings by a normalized-substring check against the reference, and short exact answers such as digit strings are never offered as candidates. The silent-audio control then tests whether the remaining clue can still replace sound. The opposite risk, offering a clue where none is needed, is handled by a \textbf{gate}: a clue is added only for detected overlap or low signal-to-noise ratio (SNR).

\subsection{Automatic Synthesis Pipeline}
\label{ssec:pipeline}
The automatic pipeline turns raw audio into AGSC, and its design follows the same principle: every clue inherits the quality of the components that build it. It diarizes speakers and detects overlap~\citep{pyannote}, marks speech regions~\citep{silero}, runs segment ASR to obtain candidate words, and estimates a coarse SNR level. For overlap plus noise, SepFormer separates the voices~\citep{sepformer}; per-speaker ASR then supplies words for shuffling (draft error $16.8\%$ on the twelve separator-selection recordings; Figure~\ref{fig:overview}b). The gate decides whether to add a clue, and the exporter records the clue with a confidence tag.

Each component must earn its place by measurement; one that lowers clue quality is dropped, not patched. We select the diarizer with lower diarization error rate (DER, $0.22$ versus $0.47$) and remove an acoustic-scene tagger with only $29\%$ accuracy, because a wrong scene label is worse than none. SNR survives only as a coarse level: the band is correct $79\%$ of the time while the point estimate has $9.9$\,dB mean error, so clues report the band and never the number. On twelve evaluation recordings, the full pipeline reaches DER $0.15$, finds the correct speaker count for $83\%$ of clips, and obtains draft character error rate (CER) $0.20$. The confidence tag makes the remaining uncertainty explicit rather than presenting an uncertain clue as ground truth.

\subsection{Supervised Training and No-Clue Evaluation}
\label{ssec:sft}
The key test is whether a post-training improvement remains after the clue text is removed. We perform SFT with LoRA on the attention projections (rank $16$, $\alpha{=}32$). For the trainable LoRA parameters $\theta$, we minimize the usual next-token loss over the serialized per-speaker target $\bar{y}$,
\begin{equation}
\mathcal{L}_{\mathrm{SFT}}(\theta)=-\sum_{t=1}^{|\bar{y}|}\log p_\theta\big(\bar{y}_t \mid \bar{y}_{<t},\,a,\,c\big).
\label{eq:sft}
\end{equation}
Equation~(\ref{eq:sft}) uses the gated AGSC clue as $c$ when the front end flags the segment and an empty context otherwise. This mixture includes both scaffolded and plain inputs. After training, the no-clue condition asks whether the behavior persists after the auxiliary text is removed. All runs use AdamW (learning rate $1.5\times10^{-4}$) for one epoch; backbone-specific execution details and shared settings appear in the supplement.

\subsection{GDPO for Gating and Streaming Outputs}
\label{ssec:gdpo}
\begin{figure*}[t]
\centering
\includegraphics[width=\textwidth]{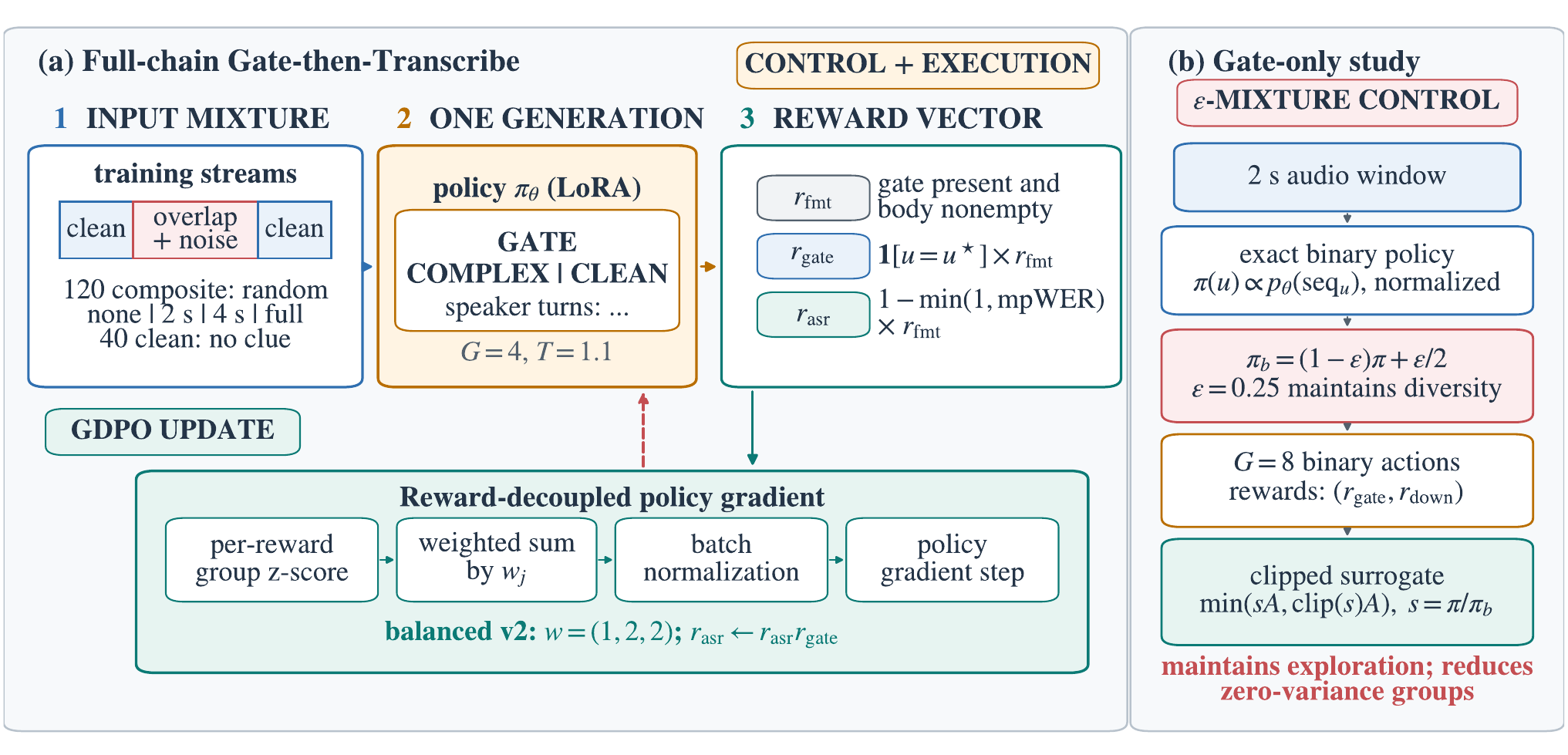}
\caption{GDPO settings. (a) Full-chain training emits one gate-plus-transcript output from a stream-like composite and a random-maturity clue. Rewards are normalized separately; balanced v2 conditions ASR credit on the correct gate. (b) Gate-only training samples from an $\varepsilon$-mixture policy $\pi_b$ around an exact target policy $\pi$ and uses $\pi/\pi_b$ in the clipped update.}
\label{fig:rlmethod}
\end{figure*}

Figure~\ref{fig:rlmethod} summarizes the full-chain and gate-only GDPO settings used below.

Output format, gate choice, and transcript quality are whole-output signals, so we train them with rewards rather than imitate a fixed answer. Group relative policy optimization (GRPO)~\citep{grpo} normalizes a scalar reward within each rollout group; with several rewards, common practice sums them first~\citep{gdpo}, which can hide different reward patterns. GDPO instead scales each reward $j$ within the rollout group $i=1,\dots,G$~\citep{gdpo},
\begin{equation}
\hat{A}_{j,i}=\frac{r_{j,i}-\mu_j}{\sigma_j+\delta},\qquad
A_i=\sum_j w_j\,\hat{A}_{j,i},
\label{eq:decouple}
\end{equation}
with $\mu_j,\sigma_j$ the group mean and standard deviation of reward $j$, $\delta=10^{-6}$, and default $w=(1,1,2)$ over (format, gate, ASR); a batch-level renormalization of $A_i$ stabilizes training. The balanced variant (v2) raises the gate weight to $w=(1,2,2)$. It also uses the conditional reward below. Writing $u$ for the gate token the model emits and $u^\star$ for the scene truth, the three rewards are
\begin{align}
r_{\mathrm{fmt}}&=\mathbb{1}[\text{gate present}\ \wedge\ \text{body nonempty}],\nonumber\\
r_{\mathrm{gate}}&=\mathbb{1}[u=u^\star]\,r_{\mathrm{fmt}},\nonumber\\
r_{\mathrm{asr}}&=\big(1-\min(1,\mathrm{mpWER})\big)\,r_{\mathrm{fmt}},
\label{eq:rewards}
\end{align}
so a malformed output earns nothing. A \emph{conditional reward} pays an easy reward only once a hard one is met,
\begin{equation}
r_{\mathrm{easy}} \leftarrow r_{\mathrm{easy}}\cdot\mathbb{1}[\,r_{\mathrm{hard}}\ge \tau\,],\quad\text{here } r_{\mathrm{asr}} \leftarrow r_{\mathrm{asr}}\cdot r_{\mathrm{gate}},
\label{eq:cond}
\end{equation}
where $\tau$ is the pass threshold of the harder reward (the binary gate reward, so $\tau=1$), preventing the policy from earning transcription reward while ignoring the gate.

The objective thus has an explicit hierarchy: a valid format exposes the decision, a correct gate opens transcription credit, and mpWER ranks the content. Separate normalization keeps each level's signal visible.

Group-relative training stalls when a confident policy samples $G$ identical actions, giving zero variance and zero gradient. For the binary gating decision we reduce this risk by computing the policy exactly,
\begin{equation}
\pi(u)=\frac{p_\theta(\mathrm{seq}_u)}{\sum_{v}p_\theta(\mathrm{seq}_v)},\quad u,v\in\{\textsc{complex},\textsc{clean}\}.
\label{eq:analytic}
\end{equation}
Here $\mathrm{seq}_u$ is the gate line for $u$; we compare whole-line probabilities because each action renders as a multi-token line. We compute the policy exactly with two forward passes per window and sample from a mixture behavior policy,
\begin{equation}
\pi_b=(1-\varepsilon)\,\pi+\varepsilon\,\mathrm{Uniform},\quad \varepsilon=0.25.
\label{eq:eps}
\end{equation}
The mixture keeps both actions possible and reduces, though does not remove, zero-variance groups. We use the importance ratio $\pi(u)/\pi_b(u)$ in a clipped proximal policy optimization (PPO)-style update to partially compensate for the behavior-policy mismatch~\citep{ppo}; this analytic policy is used only in the gate-only study, not in the full-chain rollout.

Each full-chain episode is one of $120$ three-part composites (clean, then overlap plus noise, then clean; $75\%$) or $40$ all-clean streams ($25\%$). Each episode is paired with a clue of random maturity (none, or built from the first $2$\,s, the first $4$\,s, or the full stream). In a single generation, the model emits a gate line followed by per-speaker transcription. Thus the gate token and transcript are trained jointly while the clue is available (Figure~\ref{fig:rlmethod}): the gate line is a supervised scene declaration emitted with the transcript, and clue availability is controlled by the episode's maturity schedule rather than by the emitted gate. The mpWER in Equation~(\ref{eq:rewards}) uses clean-segment pseudo-labels plus overlap gold. The gate reward uses scene truth from the composite construction, which is available without extra transcription passes; the utility-aligned signal is supplied by the gate-only study below.

In the gate-only study, a pre-measured downstream reward replaces the transcription reward,
\begin{equation}
r_{\mathrm{down}}=
\begin{cases}
\tanh(\Delta_s/0.2) & u=u^\star=\textsc{complex},\\
-0.5 & u=\textsc{complex},\,u^\star=\textsc{clean},\\
0 & u=\textsc{clean}.
\end{cases}
\label{eq:down}
\end{equation}
In Equation~(\ref{eq:down}), $\Delta_s$ is the measured mpWER benefit of injecting the clue on stream $s$ (clipped at zero), so the loop needs no extra transcription pass. Because $\Delta_s$ is a measured transcription benefit, the gate-only objective rewards clue \emph{utility}, not scene identity alone. In both reinforcement settings, Qwen3-Omni uses LoRA $(\mathrm{rank},\alpha)=(16,32)$, whereas MiniCPM-o and Ming-flash-omni use $(8,16)$. All runs use AdamW with gradient-norm clip $1$. The full chain uses learning rate $4\times10^{-5}$, $G{=}4$ rollouts, and temperature $1.1$; the gate-only study uses learning rate $2\times10^{-5}$, $G{=}8$, and PPO ratio clip $0.2$; details appear in the supplement.

\section{Experiments}
\label{sec:results}
We test three Omni models, chosen because they are strong recent open releases that differ sharply in architecture and scale, so that our conclusions do not hinge on one design: the mixture-of-experts Qwen3-Omni-30B-A3B-Instruct (30B-total, 3B-active thinker)~\citep{qwen3omni}, the dense MiniCPM-o 4.5 (9B)~\citep{minicpmo}, and the sparser Ming-flash-omni-2.0 (100B total, 6.1B active per token)~\citep{mingflash}; we write Qwen3, MiniCPM, and Ming.

CSB covers overlap, noise, meetings, streaming clue maturity, and gating. Its English speech comes from LibriSpeech and our SparseLibriMix2 manifests, two-speaker sparse mixtures built with the LibriMix recipes~\citep{librispeech,librimix}; Mandarin comes from AISHELL-3 mixes~\citep{aishell3}; noise comes from WHAM and the Music, Speech, and Noise (MUSAN) corpus~\citep{wham,musan}; and meetings come from the Augmented Multi-party Interaction (AMI) Meeting Corpus~\citep{ami}. Gold answers and predicted clues are stored separately. CSB contains $13{,}255$ training and $1{,}525$ evaluation samples. It also has dedicated splits for gate-only reinforcement ($120$ streams; $240$ paired windows), full-chain reinforcement ($160$ streams), clue maturity ($385$ prefix clues), and real-time gating ($30$ streams). The supplement gives the split table and construction details.

We report mpWER, Mandarin CER, and gating F$_1$ (the harmonic mean of precision and recall). Each hardest-third summary is selected once from the pre-training no-clue baseline, reused across conditions, and descriptive. The supervised and full-chain results cap the mpWER of each sample at $1$ before averaging, so a single degenerate output cannot dominate a mean; streaming diagnostics keep their recorded raw means. The cap is robustness, not a source of gains: capped and uncapped means agree within $1.6$ points in the supervised no-clue conditions ($3.6$ points in any condition; $0$--$3\%$ of samples exceed $1$; supplement) and within $0.6$ points on the full-chain evaluation. We report the available $95\%$ confidence intervals (CIs) for the full-set comparisons.

The experiments follow four questions that build on one another: does answer-bearing context replace listening; does scaffold-conditioned training improve transcription after the clue is removed; how does clue utility change with prefix maturity; and can reinforcement learning train the gate and full chain? The first sets up the control every later result must pass, the second tests whether quality scaffolds leave a lasting skill, the third asks when a runtime clue is worth its risk, and the fourth trains the model itself to decide when a clue should enter.

The first question is whether perception bypass actually occurs. It does, decisively (Table~\ref{tab:trap}): answer-bearing context lifts composite scores by $28$--$51$ points, yet blind-copy rates on the primary-transcript task reach $99.8\%$ under a wrong contextual answer and accuracy stays $100\%$ on silent audio, so all three models score without listening. Ming copies less often ($47\%$) mainly because it fails to follow the answer format, not because it listens better. The same control separates AGSC from this failure: with gated clues, the aggregate score on the transcription and speaker-counting probes falls to $0\%$ once audio is silenced, so whatever the clue contributes cannot replace sound.

\begin{table}[t]
\centering
{\small
\setlength{\tabcolsep}{3pt}
\begin{tabular}{lccc}
\toprule
Model & No context & +Answer & Blind-copy \\
\midrule
Qwen3 & 65.6 & 93.3 & 94.2 \\
MiniCPM  & 53.4 & 96.8 & 99.8 \\
Ming & 40.1 & 91.1 & 47.0 \\
\bottomrule
\end{tabular}
}
\caption{Perception bypass. The first two columns are composite task scores (\%); blind-copy is measured on the primary-transcript task under a wrong contextual answer. The copied answers show that the inflated score does not establish better listening.}
\label{tab:trap}
\end{table}

A target-selection case on real AMI meetings previews the intended division of labor. Without a clue, Qwen3 can transcribe speech beyond the target, consistent with target-location ambiguity; a predicted time window contains no transcript words, so it narrows where to listen while leaving word recovery to audio. This is an illustration, not a claim of uniform test-time gain.

\begin{figure}[t]
\centering
\includegraphics[width=\columnwidth]{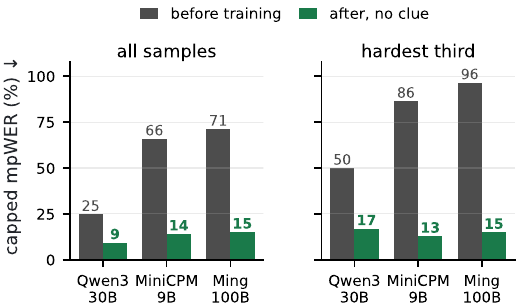}
\caption{Supervised internalization on overlap plus noise ($n{=}150$; right: the fixed, base-defined hardest $50$ of each model). Gray is pre-training; green is post-training \emph{with no clue}. Capped mean mpWER falls across all three models; under our operational definition, this no-clue improvement is internalization.}
\label{fig:intern}
\end{figure}

The second question is the payoff test: if AGSC clues direct listening rather than leak answers, training with them should leave a skill that survives their removal; a copying shortcut would vanish with the clue. It survives: on the $150$-sample overlap-plus-noise split, no-clue mpWER falls from $25\%$--$71\%$ to $9\%$--$15\%$ after AGSC fine-tuning, and from $50\%$--$96\%$ to $13\%$--$17\%$ on the hardest-third summaries (Figure~\ref{fig:intern}, Table~\ref{tab:intern}). With no clue text at test time, the improvement can only live in the weights; this is our operational evidence of internalization.

The shape of the gains supports the underused-capability hypothesis: the weaker the base, the larger the recovery (Ming $-56$, MiniCPM $-52$, Qwen3 $-16$ points), and all three converge to a narrow $9$--$15$ band. The same pattern holds for overlap without added noise ($26.6$, $64.1$, $68.3$ to $8.4$, $14.9$, $7.9$), so the effect is not tied to one test condition.

One negative result reinforces the gating principle: for MiniCPM an always-on test-time clue lags the no-clue path ($22.5$ versus $14.0$), because training clues arrived only through the gate and an unconditional clue is out of distribution; help that is not needed can hurt.

\begin{table}[b]
\centering
{\small
\setlength{\tabcolsep}{4pt}
\begin{tabular}{lccc}
\toprule
Model & No clue & Clue at test & Hardest 1/3 \\
 & (pre$\rightarrow$post) & (pre$\rightarrow$post) & (pre$\rightarrow$post) \\
\midrule
Qwen3 & 24.7$\rightarrow$\textbf{9.2} & 26.4$\rightarrow$9.1 & 49.9$\rightarrow$\textbf{16.8} \\
MiniCPM  & 65.8$\rightarrow$\textbf{14.0} & 62.6$\rightarrow$22.5 & 86.3$\rightarrow$\textbf{12.8} \\
Ming & 71.1$\rightarrow$\textbf{15.0} & 72.1$\rightarrow$14.7 & 96.4$\rightarrow$\textbf{14.9} \\
\bottomrule
\end{tabular}
}
\caption{Evaluation mpWER ($\downarrow$; per-sample cap $1$) before/after supervised training with AGSC examples ($n{=}150$). The \emph{Clue at test} column uses an unconditional test-time clue; \emph{Hardest 1/3} reports the no-clue condition and is descriptive. The $95\%$ CIs for full-set no-clue drops are $[12.1,19.0]$, $[47.9,55.7]$, and $[51.4,60.8]$ points.}
\label{tab:intern}
\end{table}

\begin{figure}[t]
\centering
\includegraphics[width=\columnwidth]{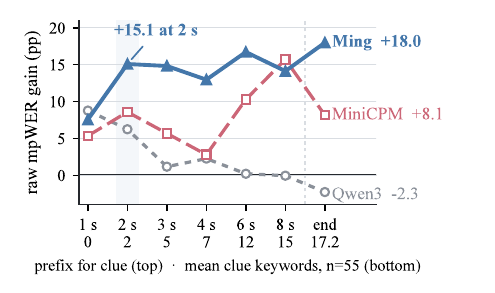}
\caption{Prefix-built clues for the pre-SFT base models on the fixed, base-defined hardest $18$ samples of each model. The model hears the full utterance; only the clue uses the indicated prefix. Curves show raw mpWER gain over no clue; lower ticks give mean clue keywords over all $55$ clips. Ming reaches $93\%$ of its terminal gain at $6$\,s.}
\label{fig:streammat}
\end{figure}

The third question moves to streaming, where a clue built from a short prefix rests on less evidence; if quality governs utility, gains should track the prefix and can even turn negative. We measure this (Figure~\ref{fig:streammat}): the full utterance is always heard, only the clue uses the prefix, and every curve reports raw mpWER gain of the pre-SFT base model on its fixed hardest $18$. Both happen: Ming reaches $15.1$ points at $2$\,s and $16.7$ at $6$\,s, or $93\%$ of its $18.0$-point terminal gain, MiniCPM peaks at $8$\,s, and for Qwen3 the terminal clue hurts this hard slice ($-2.3$ points). Utility is real but conditional, which is why injection must be selective.

Selectivity also has a second job: protecting the clean remainder of the stream. A naive time-window clue lowers Qwen3 clean-segment recall from $97.6\%$ to $73.5\%$; the instruction \emph{still transcribe the entire stream} restores it to $97.6\%$ and gives the best streaming mpWER, and an external detector saves $65\%$ of insertions by adding clues only to detected complex segments. Streaming thus needs enough prefix for a useful clue and a gate to withhold needless ones; we therefore train the gate.

The fourth question is whether reinforcement learning can train streaming control. We start with the gate alone. In the matched MiniCPM comparison, gate-only GDPO reaches the $0.875$ ceiling that $\varepsilon{=}0.25$ exploration imposes on \emph{sampled-action} accuracy under the behavior policy; evaluation decoding is greedy and not bounded by it (Table~\ref{tab:gdpores}). GRPO stops at $0.75$--$0.80$ after an early collapse (supplement). This stage learns only \emph{detection and timing}; post-training \textsc{clean} recall stays $0.80$--$0.90$ across models.

As a real-time detector, the learned gate reaches frame-level timeline F$_1$ $0.753$--$0.803$ versus $0.53$ for the external detector on the same $30$-stream split, and triggers at about $0.2$\,s versus $1.72$\,s (Table~\ref{tab:gdpores}). Gate-only training predicts \emph{whether and when} a clue would be useful; it does not train transcription. On a separate speaker-counting probe, accuracy improves by $10$ percentage points (pp) for Qwen3 and $6.2$\,pp for Ming and is unchanged for MiniCPM, suggesting the gate token trains scene perception rather than a superficial cue.

\suppressfloats[t]
\begin{table}[t]
\centering
{\small
\setlength{\tabcolsep}{2pt}
\begin{tabular}{lcccc}
\toprule
Model & \textsc{complex} & F$_1$ & Latency & Speaker \\
 & recall & (timeline) & (s) & counting \\
\midrule
Qwen3 & 0.567$\rightarrow$\textbf{1.00} & 0.727$\rightarrow$\textbf{0.803} & 0.73$\rightarrow$0.20 & 16.3$\rightarrow$\textbf{26.3} \\
MiniCPM & 0.567$\rightarrow$0.70 & 0.720$\rightarrow$0.769 & 0.53$\rightarrow$0.17 & 35.0$\rightarrow$35.0 \\
Ming & 0.20$\rightarrow$\textbf{1.00} & 0.522$\rightarrow$\textbf{0.753} & 1.19$\rightarrow$0.23 & 15.0$\rightarrow$\textbf{21.2} \\
\bottomrule
\end{tabular}
}
\caption{Gate-only GDPO (pre$\rightarrow$post). \textsc{Complex} recall uses $30$ complex windows; F$_1$/latency use $30$ streams; speaker counting uses a separate $80$-example probe. External F$_1$/latency are $0.53/1.72$\,s; Table~\ref{tab:chain} gives full-chain mpWER.}
\label{tab:gdpores}
\end{table}

The full-chain study optimizes one gate-plus-speaker-attributed-transcript output. It uses streaming clues from only $160$ streams, versus $3{,}500$ noise-stage examples in the supervised runs. Table~\ref{tab:chain} reports mpWER and, from the same runs, no-clue complex-gate recall, so each row is judged on both outputs. With no clue at test time, mpWER drops by $14.1$ points for MiniCPM (v1) and $64.2$ for Ming; the reported CIs for both changes exclude zero. The Qwen3 drop is $2.8$ points and its reported CI $[-1.0,7.2]$ includes zero; Qwen3 starts strongest ($33.0$) and has the least headroom. Ming starts at the $100.0$ boundary with no scorable transcript lines, so its large gain includes restoring a valid output format on top of improved listening.

MiniCPM exposes an explicit trade-off, reported from both sides. Under a dominant transcription weight (v1), no-clue mpWER improves most ($58.5\rightarrow44.4$) but no-clue complex-gate recall falls ($0.49\rightarrow0.38$). Balanced v2 sets $w=(1,2,2)$ in Equation~(\ref{eq:decouple}) and conditions transcription credit on a correct gate using Equation~(\ref{eq:cond}): recall recovers above baseline ($0.78$) with a smaller transcription gain ($55.5$). The two variants are separate reward configurations trained independently, not checkpoints selected from one run; a deployment picks the adapter by which output matters more.

\begin{table}[t]
\centering
{\small
\setlength{\tabcolsep}{4pt}
\begin{tabular}{lccccc}
\toprule
Model & No clue & 2\,s & 4\,s & Full & Gate \\
\midrule
Qwen3 base & 33.0 & 31.9 & 33.7 & 31.6 & 0.82 \\
Qwen3 chain & \textbf{30.2} & 29.6 & \textbf{27.9} & 29.9 & 0.87 \\
MiniCPM base & 58.5 & 56.1 & 58.7 & 57.0 & 0.49 \\
MiniCPM chain v1 & \textbf{44.4} & 49.6 & 50.9 & \textbf{42.9} & 0.38 \\
MiniCPM chain v2 & 55.5 & 50.4 & 47.0 & 48.6 & \textbf{0.78} \\
Ming base & 100.0 & 80.6 & 83.7 & 89.9 & 0.82 \\
Ming chain & \textbf{35.8} & 33.6 & 32.2 & 38.4 & \textbf{0.98} \\
\bottomrule
\end{tabular}
}
\caption{Full-chain capped mean mpWER ($\downarrow$; per-sample cap $1$; $n{=}55$ per condition/model; capping shifts these means by at most $0.6$ points). \emph{Gate} is complex-gate recall in the no-clue condition from the same runs. No-clue drop $95\%$ CIs: Qwen3 $[-1.0,7.2]$, MiniCPM v1 $[7.1,21.9]$, Ming $[57.1,70.9]$.}
\label{tab:chain}
\end{table}

Heavy training can cause \textbf{gate collapse}: the model reports \textsc{complex} for almost every clean stream (accuracy $0$, $0$, and $0.11$). A forced-choice probe shows a median complex probability of $1.00$ for both stream types, so no soft threshold can help. At this extreme, a logit gap near $20$ drives both the per-token gradient $p(1{-}p)$, with $p$ the gate probability, and the group advantage toward zero, so a short supervised step provides the missing gradient. After $90$ steps, clean-stream gate accuracy returns to $1.0$ for all models and macro-averaged mixed-stream mpWER also decreases (supplement); the runs in Table~\ref{tab:chain} stop before this regime, and the collapse appears only under heavier training.

Scaffolding is practical only if it stays cheap. AGSC adds $113$--$157$ prompt tokens, but audio processing dominates inference. Wall-clock overhead is $0.03$--$0.21$\,s ($2.9$--$11.9\%$) except for Ming on overlap plus noise, where task completion produces an apparent $69\%$ increase; see the supplement for the protocol. Clues are built offline in $1.56$\,s; caching removes repeat cost.

Two usage modes follow: a streaming mode that pairs the learned gate (timeline F$_1$ $0.75$--$0.80$, $\sim$$0.2$\,s trigger latency) with prefix-built clues, paying overhead only when it fires, and a no-clue mode using the plain audio prompt. Gate metrics judge the first mode and no-clue mpWER the second.

\section{Discussion}
Our interpretation is that pretraining already equips Omni models with the acoustic skills complex scenes demand: during SFT a clue acts as a temporary trigger that engages those skills on hard segments; once training ends the trigger is no longer needed, and this no-clue persistence is internalization. The worry that the models merely copy clue text is ruled out by design and results: the answer-overlap screen removes copyable strings, the silent control zeroes aggregate scores, and after training an added clue adds no meaningful gain over no clue (Table~\ref{tab:intern}).

Each check answers a different question, whether a clue helps (context on), can replace audio (silence), persists after removal (no clue), or is timed correctly (streaming); internalization is judged against each pretrained baseline. Clue utility depends on the scene and the clue-building prefix (Figure~\ref{fig:streammat}): a wordless window narrows target location, more prefix audio does not always help, so templates match clue type to ambiguity and the gate limits entry. The results thus separate what persists without help (no-clue mpWER), when a clue enters (gate metrics), and joint accuracy (full-chain mpWER).

\section{Conclusion}
This paper turns a tempting shortcut into a design principle. Answer-bearing context lets a model score without listening; we call this perception bypass and expose it by silencing audio. It anchors three contributions: AGSC scaffolds guide listening while limiting copying; CSB adds explicit shortcut controls; and internalization is judged by gains that persist after clue removal. After SFT, no-clue capped mpWER on overlap plus noise falls from $25\%$--$71\%$ to $9\%$--$15\%$, our evidence of internalization. For streaming, gate-only GDPO learns when a clue is worth using; full-chain GDPO trains gate and transcript jointly. Future work extends scene-matched clues and stresses the gate under domain shift. The deeper principle is simple: context should \mbox{scaffold perception, not replace it.}

\bibliography{refs}

@misc{qwen3omni,
  author={Xu, Jin and Guo, Zhifang and Hu, Hangrui and Chu, Yunfei and Wang, Xiong and others},
  title={Qwen3-Omni Technical Report},
  year={2025},
  eprint={2509.17765},
  archivePrefix={arXiv},
  primaryClass={cs.CL},
}

@misc{minicpmo,
  author={Cui, Junbo and Xu, Bokai and Wang, Chongyi and Yu, Tianyu and Sun, Weiyue and others},
  title={MiniCPM-o 4.5: Towards Real-Time Full-Duplex Omni-Modal Interaction},
  year={2026},
  eprint={2604.27393},
  archivePrefix={arXiv},
  primaryClass={cs.CL},
}

@inproceedings{pyannote,
  author={Bredin, Herv\'{e}},
  title={pyannote.audio 2.1 speaker diarization pipeline: principle, benchmark, and recipe},
  booktitle={Proceedings of the Annual Conference of the International Speech Communication Association (INTERSPEECH)},
  year={2023},
  pages={1983--1987},
  doi={10.21437/Interspeech.2023-105},
}

@inproceedings{sepformer,
  author={Subakan, Cem and Ravanelli, Mirco and Cornell, Samuele and Bronzi, Mirko and Zhong, Jianyuan},
  title={Attention Is All You Need in Speech Separation},
  booktitle={Proceedings of the IEEE International Conference on Acoustics, Speech and Signal Processing (ICASSP)},
  year={2021},
  pages={21--25},
  doi={10.1109/ICASSP39728.2021.9413901},
}

@inproceedings{whisper,
  author={Radford, Alec and Kim, Jong Wook and Xu, Tao and Brockman, Greg and McLeavey, Christine and Sutskever, Ilya},
  title={Robust Speech Recognition via Large-Scale Weak Supervision},
  booktitle={Proceedings of the International Conference on Machine Learning (ICML)},
  volume={202},
  publisher={PMLR},
  year={2023},
  pages={28492--28518},
}

@misc{silero,
  author={{Silero Team}},
  title={Silero {VAD}: pre-trained enterprise-grade Voice Activity Detector ({VAD}), Number Detector and Language Classifier},
  year={2024},
  publisher={GitHub},
  howpublished={\url{https://github.com/snakers4/silero-vad}},
}

@inproceedings{lora,
  author={Hu, Edward J. and Shen, Yelong and Wallis, Phillip and Allen-Zhu, Zeyuan and Li, Yuanzhi and Wang, Shean and Wang, Lu and Chen, Weizhu},
  title={LoRA: Low-Rank Adaptation of Large Language Models},
  booktitle={Proceedings of the International Conference on Learning Representations (ICLR)},
  year={2022},
}

@misc{grpo,
  author={Shao, Zhihong and Wang, Peiyi and Zhu, Qihao and Xu, Runxin and Song, Junxiao and others},
  title={DeepSeekMath: Pushing the Limits of Mathematical Reasoning in Open Language Models},
  year={2024},
  eprint={2402.03300},
  archivePrefix={arXiv},
  primaryClass={cs.CL},
}

@misc{ppo,
  author={Schulman, John and Wolski, Filip and Dhariwal, Prafulla and Radford, Alec and Klimov, Oleg},
  title={Proximal Policy Optimization Algorithms},
  year={2017},
  eprint={1707.06347},
  archivePrefix={arXiv},
  primaryClass={cs.LG},
}

@inproceedings{gdpo,
  author={Liu, Shih-Yang and Dong, Xin and Lu, Ximing and Diao, Shizhe and Belcak, Peter and Liu, Mingjie and Chen, Min-Hung and Yin, Hongxu and Wang, Yu-Chiang Frank and Cheng, Kwang-Ting and Choi, Yejin and Kautz, Jan and Molchanov, Pavlo},
  title={{GDPO}: Group reward-Decoupled Normalization Policy Optimization for Multi-reward {RL} Optimization},
  booktitle={Proceedings of the International Conference on Machine Learning (ICML)},
  year={2026},
  url={https://icml.cc/virtual/2026/poster/63333},
}

@inproceedings{cpwer,
  author={Watanabe, Shinji and Mandel, Michael and Barker, Jon and Vincent, Emmanuel and others},
  title={{CHiME-6} Challenge: Tackling Multispeaker Speech Recognition for Unsegmented Recordings},
  booktitle={Proceedings of the International Workshop on Speech Processing in Everyday Environments (CHiME)},
  year={2020},
  pages={1--7},
  doi={10.21437/CHiME.2020-1},
}

@inproceedings{ami,
  author={Carletta, Jean and Ashby, Simone and Bourban, Sebastien and Flynn, Mike and Guillemot, Mael and Hain, Thomas and Kadlec, Jaroslav and Karaiskos, Vasilis and Kraaij, Wessel and Kronenthal, Melissa and Lathoud, Guillaume and Lincoln, Mike and Lisowska, Agnes and McCowan, Iain and Post, Wilfried and Reidsma, Dennis and Wellner, Pierre},
  title={The {AMI} Meeting Corpus: A Pre-announcement},
  booktitle={Machine Learning for Multimodal Interaction, Second International Workshop},
  series={Lecture Notes in Computer Science},
  volume={3869},
  year={2006},
  pages={28--39},
  publisher={Springer},
  doi={10.1007/11677482_3},
}

@inproceedings{librispeech,
  author={Panayotov, Vassil and Chen, Guoguo and Povey, Daniel and Khudanpur, Sanjeev},
  title={{LibriSpeech}: An {ASR} Corpus Based on Public Domain Audio Books},
  booktitle={Proceedings of the IEEE International Conference on Acoustics, Speech and Signal Processing (ICASSP)},
  year={2015},
  pages={5206--5210},
  doi={10.1109/ICASSP.2015.7178964},
}

@inproceedings{wham,
  author={Wichern, Gordon and Antognini, Joe and Flynn, Michael and Zhu, Licheng Richard and McQuinn, Emmett and Crow, Dwight and Manilow, Ethan and Le Roux, Jonathan},
  title={{WHAM!}: Extending Speech Separation to Noisy Environments},
  booktitle={Proceedings of the Annual Conference of the International Speech Communication Association (INTERSPEECH)},
  year={2019},
  pages={1368--1372},
  doi={10.21437/Interspeech.2019-2821},
}

@inproceedings{aishell3,
  author={Shi, Yao and Bu, Hui and Xu, Xin and Zhang, Shaoji and Li, Ming},
  title={{AISHELL-3}: A Multi-Speaker Mandarin {TTS} Corpus},
  booktitle={Proceedings of the Annual Conference of the International Speech Communication Association (INTERSPEECH)},
  year={2021},
  pages={2756--2760},
  doi={10.21437/Interspeech.2021-755},
}

@misc{musan,
  author={Snyder, David and Chen, Guoguo and Povey, Daniel},
  title={{MUSAN}: A Music, Speech, and Noise Corpus},
  year={2015},
  eprint={1510.08484},
  archivePrefix={arXiv},
  primaryClass={cs.SD},
}

@inproceedings{rag,
  author={Lewis, Patrick and Perez, Ethan and Piktus, Aleksandra and others},
  title={Retrieval-Augmented Generation for Knowledge-Intensive {NLP} Tasks},
  booktitle={Advances in Neural Information Processing Systems (NeurIPS)},
  volume={33},
  year={2020},
  pages={9459--9474},
}

@inproceedings{cot,
  author={Wei, Jason and Wang, Xuezhi and Schuurmans, Dale and others},
  title={Chain-of-Thought Prompting Elicits Reasoning in Large Language Models},
  booktitle={Advances in Neural Information Processing Systems (NeurIPS)},
  volume={35},
  year={2022},
  pages={24824--24837},
  doi={10.52202/068431-1800},
}

@inproceedings{salmonn,
  author={Tang, Changli and Yu, Wenyi and Sun, Guangzhi and Chen, Xianzhao and others},
  title={{SALMONN}: Towards Generic Hearing Abilities for Large Language Models},
  booktitle={Proceedings of the International Conference on Learning Representations (ICLR)},
  volume={2024},
  year={2024},
  pages={16607--16629},
}

@inproceedings{audiocontrib,
  author={He, Haolin and Du, Xingjian and Sun, Renhe and Dai, Zheqi and Xiao, Yujia and Yang, Mingru and Zhou, Jiayi and Li, Xiquan and Liu, Zhengxi and Liang, Zining and Wu, Chunyat and He, Qianhua and Lee, Tan and Chen, Xie and Zheng, Wei-Long and Wang, Weiqiang and Plumbley, Mark D. and Liu, Jian and Kong, Qiuqiang},
  title={Measuring Audio's Impact on Correctness: Audio-Contribution-Aware Post-Training of Large Audio Language Models},
  booktitle={Proceedings of the International Conference on Learning Representations (ICLR)},
  year={2026},
  url={https://iclr.cc/virtual/2026/poster/10007069},
}

@misc{iraf,
  author={Zhong, Tao and Deng, Jiajun and Kuzmin, Nikita and Zhu, Yinke and Cao, Tianxiang and Tsoi, Tristan and Tan, Zhili and Lui, Simon and Liu, Xunying},
  title={{IRAF}: Interference-Resilient Adaptive Fusion for Noise-Robust End-to-End Full-Duplex Spoken Dialogue Systems},
  year={2026},
  eprint={2606.06559},
  archivePrefix={arXiv},
  primaryClass={cs.SD},
}

@misc{qwenaudio,
  author={Chu, Yunfei and Xu, Jin and Zhou, Xiaohuan and Yang, Qian and Zhang, Shiliang and Yan, Zhijie and Zhou, Chang and Zhou, Jingren},
  title={Qwen-Audio: Advancing Universal Audio Understanding via Unified Large-Scale Audio-Language Models},
  year={2023},
  eprint={2311.07919},
  archivePrefix={arXiv},
  primaryClass={eess.AS},
}

@inproceedings{tsot,
  author={Kanda, Naoyuki and Gaur, Yashesh and Wang, Xiaofei and Meng, Zhong and Yoshioka, Takuya},
  title={Serialized Output Training for End-to-End Overlapped Speech Recognition},
  booktitle={Proceedings of the Annual Conference of the International Speech Communication Association (INTERSPEECH)},
  year={2020},
  pages={2797--2801},
  doi={10.21437/Interspeech.2020-999},
}

@inproceedings{tsvad,
  author={Medennikov, Ivan and Korenevsky, Maxim and Prisyach, Tatiana and Khokhlov, Yuri and Korenevskaya, Mariya and Sorokin, Ivan and Timofeeva, Tatiana and Mitrofanov, Anton and Andrusenko, Andrei and Podluzhny, Ivan and Laptev, Aleksandr and Romanenko, Aleksei},
  title={Target-Speaker Voice Activity Detection: A Novel Approach for Multi-Speaker Diarization in a Dinner Party Scenario},
  booktitle={Proceedings of the Annual Conference of the International Speech Communication Association (INTERSPEECH)},
  year={2020},
  pages={274--278},
  doi={10.21437/Interspeech.2020-1602},
}

@inproceedings{chime8,
  author={Cornell, Samuele and Park, Tae Jin and Huang, He and Boeddeker, Christoph and Chang, Xuankai and Maciejewski, Matthew and Wiesner, Matthew S. and Garcia, Paola and Watanabe, Shinji},
  title={The {CHiME-8} {DASR} Challenge for Generalizable and Array Agnostic Distant Automatic Speech Recognition and Diarization},
  booktitle={Proceedings of the International Workshop on Speech Processing in Everyday Environments (CHiME)},
  year={2024},
  pages={1--6},
  doi={10.21437/CHiME.2024-1},
}

@inproceedings{clas,
  author={Pundak, Golan and Sainath, Tara N. and Prabhavalkar, Rohit and Kannan, Anjuli and Zhao, Ding},
  title={Deep Context: End-to-End Contextual Speech Recognition},
  booktitle={Proceedings of the IEEE Spoken Language Technology Workshop (SLT)},
  year={2018},
  pages={418--425},
  doi={10.1109/SLT.2018.8639034},
}

@inproceedings{diarizationlm,
  author={Wang, Quan and Huang, Yiling and Zhao, Guanlong and Clark, Evan and Xia, Wei and Liao, Hank},
  title={DiarizationLM: Speaker Diarization Post-Processing with Large Language Models},
  booktitle={Proceedings of the Annual Conference of the International Speech Communication Association (INTERSPEECH)},
  year={2024},
  pages={3754--3758},
  doi={10.21437/Interspeech.2024-209},
}

@inproceedings{hyporadise,
  author={Chen, Chen and Hu, Yuchen and Yang, Chao-Han Huck and Siniscalchi, Sabato Marco and Chen, Pin-Yu and Chng, Eng-Siong},
  title={HyPoradise: An Open Baseline for Generative Speech Recognition with Large Language Models},
  booktitle={Advances in Neural Information Processing Systems},
  note={Datasets and Benchmarks Track},
  volume={36},
  year={2023},
  pages={31665--31688},
  doi={10.52202/075280-1375},
}

@article{shortcut,
  author={Geirhos, Robert and Jacobsen, J{\"o}rn-Henrik and Michaelis, Claudio and Zemel, Richard and Brendel, Wieland and Bethge, Matthias and Wichmann, Felix A.},
  title={Shortcut Learning in Deep Neural Networks},
  journal={Nature Machine Intelligence},
  volume={2},
  number={11},
  year={2020},
  pages={665--673},
  doi={10.1038/s42256-020-00257-z},
  note={DOI: 10.1038/s42256-020-00257-z},
}

@inproceedings{vqa2,
  author={Goyal, Yash and Khot, Tejas and Summers-Stay, Douglas and Batra, Dhruv and Parikh, Devi},
  title={Making the V in VQA Matter: Elevating the Role of Image Understanding in Visual Question Answering},
  booktitle={Proceedings of the IEEE Conference on Computer Vision and Pattern Recognition (CVPR)},
  year={2017},
  pages={6904--6913},
  doi={10.1109/CVPR.2017.670},
}

@misc{ctxdistill,
  author={Snell, Charlie and Klein, Dan and Zhong, Ruiqi},
  title={Learning by Distilling Context},
  year={2022},
  eprint={2209.15189},
  archivePrefix={arXiv},
  primaryClass={cs.CL},
}

@inproceedings{mmau,
  author={Sakshi, Sakshi and Tyagi, Utkarsh and Kumar, Sonal and Seth, Ashish and Selvakumar, Ramaneswaran and Nieto, Oriol and Duraiswami, Ramani and Ghosh, Sreyan and Manocha, Dinesh},
  title={{MMAU}: A Massive Multi-Task Audio Understanding and Reasoning Benchmark},
  booktitle={Proceedings of the International Conference on Learning Representations (ICLR)},
  volume={2025},
  year={2025},
  pages={84929--84964},
  url={https://proceedings.iclr.cc/paper_files/paper/2025/file/d36f208919582785db965fe648b9fe59-Paper-Conference.pdf},
}

@inproceedings{airbench,
  author={Yang, Qian and Xu, Jin and Liu, Wenrui and Chu, Yunfei and Jiang, Ziyue and Zhou, Xiaohuan and Leng, Yichong and Lv, Yuanjun and Zhao, Zhou and Zhou, Chang and Zhou, Jingren},
  title={AIR-Bench: Benchmarking Large Audio-Language Models via Generative Comprehension},
  booktitle={Proceedings of the Annual Meeting of the Association for Computational Linguistics (ACL)},
  year={2024},
  pages={1979--1998},
  doi={10.18653/v1/2024.acl-long.109},
  url={https://aclanthology.org/2024.acl-long.109/},
}

@inproceedings{dynamicsuperb,
  author={Huang, Chien-yu and Lu, Ke-Han and Wang, Shih-Heng and Hsiao, Chi-Yuan and Kuan, Chun-Yi and Wu, Haibin and Arora, Siddhant and Chang, Kai-Wei and Shi, Jiatong and Peng, Yifan and Sharma, Roshan and Watanabe, Shinji and Ramakrishnan, Bhiksha and Shehata, Shady and Lee, Hung-yi},
  title={Dynamic-SUPERB: Towards a Dynamic, Collaborative, and Comprehensive Instruction-Tuning Benchmark for Speech},
  booktitle={Proceedings of the IEEE International Conference on Acoustics, Speech and Signal Processing (ICASSP)},
  year={2024},
  pages={12136--12140},
  doi={10.1109/ICASSP48485.2024.10448257},
}

@inproceedings{blindfaith,
  author={Deng, Ailin and Cao, Tri and Chen, Zhirui and Hooi, Bryan},
  title={Words or Vision: Do Vision-Language Models Have Blind Faith in Text?},
  booktitle={Proceedings of the IEEE/CVF Conference on Computer Vision and Pattern Recognition (CVPR)},
  year={2025},
  pages={3867--3876},
  doi={10.1109/CVPR52734.2025.00366},
}

@inproceedings{cmm,
  author={Leng, Sicong and Xing, Yun and Cheng, Zesen and Zhou, Yang and Zhang, Hang and Li, Xin and Zhao, Deli and Lu, Shijian and Miao, Chunyan and Bing, Lidong},
  title={The Curse of Multi-Modalities: Evaluating Hallucinations of Large Multimodal Models across Language, Visual, and Audio},
  booktitle={Advances in Neural Information Processing Systems (NeurIPS)},
  volume={38},
  year={2025},
}

@misc{audiohalluc,
  author={Nishimura, Taichi and Nakada, Shota and Kondo, Masayoshi},
  title={On the Audio Hallucinations in Large Audio-Video Language Models},
  year={2024},
  eprint={2401.09774},
  archivePrefix={arXiv},
  primaryClass={cs.MM}
}

@misc{mingflash,
  author={{Inclusion AI}},
  title={Ming-Flash-Omni: A Sparse, Unified Architecture for Multimodal Perception and Generation},
  year={2025},
  eprint={2510.24821},
  archivePrefix={arXiv},
  primaryClass={cs.CV},
}

@misc{librimix,
  author={Cosentino, Joris and Pariente, Manuel and Cornell, Samuele and Deleforge, Antoine and Vincent, Emmanuel},
  title={LibriMix: An Open-Source Dataset for Generalizable Speech Separation},
  year={2020},
  eprint={2005.11262},
  archivePrefix={arXiv},
  primaryClass={eess.AS},
}

@article{wavlm,
  author={Chen, Sanyuan and Wang, Chengyi and Chen, Zhengyang and Wu, Yu and Liu, Shujie and Chen, Zhuo and Li, Jinyu and Kanda, Naoyuki and Yoshioka, Takuya and Xiao, Xiong and Wu, Jian and Zhou, Long and Ren, Shuo and Qian, Yanmin and Qian, Yao and Wu, Jian and Zeng, Michael and Yu, Xiangzhan and Wei, Furu},
  title={{WavLM}: Large-Scale Self-Supervised Pre-Training for Full Stack Speech Processing},
  journal={IEEE Journal of Selected Topics in Signal Processing},
  volume={16},
  number={6},
  pages={1505--1518},
  year={2022},
  doi={10.1109/JSTSP.2022.3188113},
  note={DOI: 10.1109/JSTSP.2022.3188113},
}

\end{document}